\documentstyle[12pt,aaspp4,epsfig]{article}

\def\beq{\begin{equation}}
\def\enq{\end{equation}}
\def\bea{\begin{array}}
\def\ena{\end{array}}
\def\Mesz{M\'esz\'aros~}

\newcommand{\siml}{\lower4pt \hbox{$\buildrel < \over \sim$}}
\newcommand{\simg}{\lower4pt \hbox{$\buildrel > \over \sim$}}

\begin{document}

\title{Detectability of GRB Iron Lines by Swift, Chandra and XMM}
\author{ L. J. Gou$^1$,  P. M\'{e}sz\'{a}ros$^{1,2,3}$ and T. R. Kallman$^4$}

\noindent
$^1$Dpt. Astron. \& Astrophys., Pennsylvania State University, University Park, PA 16802\\
$^2$Dpt. of Physics, Pennsylvania State University, University Park, PA 16802\\
$^3$The institute for Advanced Study, Princeton, NJ 08540\\
$^4$NASA Goddard Space Flight Center, LHEA, Code 665, Greenbelt, MD 20771\\

\begin{center} Revised :~~{01/05/2005} \end{center}

\begin{abstract}

The rapid acquisition of positions by the upcoming Swift satellite
will allow the monitoring for X-ray lines in GRB afterglows at
much earlier epochs than was previously feasible. We calculate the
possible significance levels of iron line detections as a function
of source redshift and observing time after the trigger, for the
Swift XRT, Chandra ACIS and XMM Epic detectors. For bursts with
standard luminosities, decay rates and equivalent widths of 1 keV
assumed constant starting at early source-frame epochs, Swift may be
able to detect lines up to $z \sim 1.5$ with a significance of $\simg
3\sigma$ for times $t\siml 10^4$ s. The same lines would be detectable
with $\simg 4\sigma$ significance at $z\siml 6$ by Chandra, and at
$z\siml 8$ by XMM, for times of $t\siml 10^5$ s. For similar bursts
with a variable equivalent width which peaks at 1 keV between 0.5 and 1
days in the source frame, Swift achieves the same significance level
for $z\sim 1$ at $t\sim 1$ day, while Chandra reaches the previous
detection significances around $t\sim 1-2$ days for $z\sim 2-4$, i.e.
the line is detectable near the peak equivalent width times, and
undetectable at earlier or later times. For afterglows in the upper
range of initial X-ray luminosites afterglows, which may also be typical
of pop. III bursts, similar significance levels are obtained out to
substantially higher redshifts. A distinction between broad and
narrow lines to better than $3\sigma$ is possible with Chandra and
XMM out to $z\sim 2$ and $\sim 6.5$, respectively, while Swift can
do so up to $z\sim 1$, for standard burst parameters. A
distinction between different energy centroid lines of 6.4 keV vs.
6.7 KeV (or 6.7 keV vs. Cobalt 7.2 keV) is possible up to $z\siml
0.6$, $1.2$, and $2$ ( $z\siml 1, \ 5, \ 7.5$), with Swift,
Chandra, and XMM respectively. For the higher luminosity bursts,
Swift is able to distinguish at the $5\sigma$ level between a
broad and a narrow line out to $z\siml 5$, and between a 6.7 keV
vs. a 7.2 keV line center out to $z\siml 5$ for times of  $t\siml
10^4$ s.

\end{abstract}

\section{Introduction}
\label{sec:intro}

The detection of Fe K-$\alpha$ X-ray lines can play an important
role in understanding the nature of GRB. It may provide insights
into the nature and details of the GRB progenitor, e.g. through
possible differences in the line properties of short GRBs and
long GRBs. While long GRBs are convincingly associated with the
collapse of massive stars, short GRBs are, lacking other evidence,
largely believed to arise from compact star mergers such as NS/NS
or NS/BH systems. Thus in short bursts one might expect an ambient
gas density which is lower, and a stellar remnant or stellar funnel
which is more compact than in the long burst case. In the latter,
the explosion may be expected to take place in a higher density
medium, e.g. a star forming region, or the pre-burst wind of the
progenitor, and the progenitor is spatially more extended.

In the case of long GRBs, several mechanisms have been proposed
for generating iron lines. These fall mainly into two categories:
distant reprocessor or nearby reprocessor models. Both of them
assume photoionization and reprocessing by a stellar remnant
medium outside the source of continuum photons associated with the
afterglow. In the distant models, based on the supranova paradigm,
the x-rays from the burst and early  afterglow emission illuminate
iron-rich material at a distance of $\simg 10^{16} \ cm$ which is
outside the fireball region, deposited there by a supernova
explosion occurring months before the GRB event (Lazzati et al.
1999). In this case, the line intensity variations are thought to
be due to light travel time effects between the GRB and the
reprocessor. Alternatively, in the nearby reprocessor models, the
line emission is attributed to the interaction of a long lasting
outflow from the central engine with the progenitor stellar
envelope at distances $R \lesssim 10^{13} cm$ (Rees \& \Mesz 2000;
\Mesz \& Rees 2001; B\"{o}ttcher \& Fryer 2001).  In this case,
line variations are attributed to changes or decay of the
photoionizing radiation continuum or jet. Therefore, iron emission
line feature (e.g. equivalent width ($EW$)) produced by the
different mechanisms will be different (Ballantyne, et al. 2002,
Kallman et al., 2002).

Among the well-localized GRBs, $\sim$ 90\% have x-ray afterglows
and about 60\% of bursts with x-ray afterglow detections are also
detected in the optical band. However the other 40\% are optically
dark. In several cases (e.g. GRB 970508, GRB 970828, GRB 991216,
GRB990705), the x-ray redshifts derived from the iron lines were
consistent with those from optical spectroscopy of the host galaxy
(Piro 2002). This shows that measurements of the redshift by X-ray
spectroscopy is more than mere a possibility, and can provide reliable
results, which is particularly interesting when the optical
spectroscopy is difficult or non-existent, as in dark bursts.

The possible role of iron lines in tracing the high redshift
Universe has been discussed by \Mesz \& Rees (2003), Ghisellini
et al (1999) and others. Since little absorption is expected
from either our galaxy or the intergalactic medium in the X-ray
band above $\sim 0.2$ keV, one can expect the Fe K-$\alpha$ line
to be in principle detectable up to redshifts 30, if GRB are
present there and if the signal to noise ratio is sufficient
for a given spectrometric instrument. It is this latter question
which we address here.

So far, 5 iron line features have been detected: GRB 970508 (Piro
et al. 1999, BeppoSAX), GRB 970828 (Yoshida et al. 1999, ASCA),
GRB 990705 (Amati et al. 2000, BeppoSAX, prompt x-ray emission),
GRB 991216 (Piro et al. 2000,  Chandra), GRB 000214 (Antonelli et
al. 2000, BeppoSAX). However, almost all of them are detected
marginally. A compilation of the significance levels is GRB
970508: $\sim 2.46 \sigma$ (99.3\%); GRB 970828:$\sim 2.12 \sigma$
(98.3\%); GRB 990705: $\sim 1.71 \sigma$ (95.6\%) (based on the
fit results of Amati et al. (2000) and our own); GRB 991216, Piro
et al.: $\sim 2.58 \sigma$ (99.5\%); Ballantyne et al.: $\sim 2.06
\sigma$ (98\%); GRB 000214: 3$\sigma$. Because of the low signal
to noise in these previous observations, it is difficult to
differentiate between the two main classes of line production
models. One example here is GRB 991216, for which Piro et al.
(2000) argued that it can be explained well by a distant
reprocessor model, whereas Ballantyne et al. (2002) argued that it
could be explained in a reflected emission model too, which is
compatible with the nearby model. Thus, the detection of higher
signal to noise line features is necessary. This would be
necessary also in order to have some confidence in the utility of
the lines as redshift tracers.

In a number of other bursts, several low Z (ionic charge) lines
have been reported with XMM (Reeves et al. 2002; Watson et al.
2002, 2003; see also Table 1 in O'Brian et al. (2003) for the
summary of all detected low Z x-ray lines). The non-detection of
iron lines at a significant level in these same objects by
XMM, and the non-detection of significant lower Z elements in the
objects where Chandra detected Fe lines, is an interesting
question which remains to be clarified. Observations at an earlier
phase of the afterglow, when the lines may be easier to detect,
could throw light on this question. Swift, due for launch in late
2004, has spectral capabilities and a very short slewing time
($t\lesssim\ 1$ minute). Thus, if strong enough lines are produced
at minutes to hours, as many $\sim 100 \ {\rm yr^{-1}}$ bursts
with lines might be detected with higher signal to noise ratio
than heretofore. On the other hand, the significance level of most
of the GRB afterglow line systems reported in the literature have
been put into question (Sako, Harrison \& Rutledge 2004). This
underlines the uncertainties of the previous detections, and the
importance of finding, or not finding, such lines with the
improved detection sensitivities  made possible by earlier
spectral measurements when Swift comes on line. In this paper, we
investigate the detectability of Iron line emission, both with
Swift and with Chandra and XMM, and address the question of how
far can bursts be detected and their redshifts measured with a
quantifiable confidence level ($\simg 3 \sigma$), as function of
the epoch after the trigger when the line forms and is observed.

\section{Model and Procedure}
\label{sec:model}

The start time of the GRB X-ray afterglow depends on various
details about the external density and the parameters of the
burst. Here we adopt the usual phenomenological definition which
takes the start of the X-ray afterglow $t_i$ to coincide with the
end of GRB itself (i.e. of its $\gamma$-ray duration).

For the X-ray afterglow continuum flux, we use the observer frame
flux $F_E(E,t)$ as a function of observed energy $E$ and observer
time $t$ for a spectrum parameterized as $F_E\propto E^at^b$
(Lamb \& Reichart 2000), generalized to a double power law to
account for the jet break and consequent steepening of the light
curve. The source is assumed to have an initial luminosity
$L_{E,t_i}$ which in the rest frame is constant between the trigger
time and source frame prompt phase duration time $T/(1+z)$, which
we take nominally to be 20 s. This is followed by an initial power
law decay $\propto t^{b1}$, and after a time $t_{br}/(1+z)$ (the
source frame jet break time, nominally taken to be 0.5 days), the
decay is assumed to follow a steeper power law $\propto t^{b2}$,
due to the jet having a finite size. The observer-frame spectral
flux is then
\begin{equation}
F_E(z,t)= {L_{E,t_i} \over 4\pi D_l(z)^2 (1+z)^{-1-a+{b1}}}
\Bigl[ ({t \over t_i})^{b1} H({t_{br}\over t}) +
   ({t_{br} \over t_i})^{b1} ({t \over t_{br}})^{b2} H({t\over t_{br}})\Bigr]~,
\label{eq:Fxdouble}
\end{equation}
Here $L_{E,t_i}= 10^{49.3} L_{x,50} E_{\rm keV}^{a}\
{\rm ergs\ s^{-1}\ keV^{-1}}$ is the initial GRB afterglow
isotropic-equivalent luminosity per energy corresponding to an X-ray
luminosity in the 0.2-10 keV band of $L_x\simeq 10^{50}$ erg s$^{-1}$,
assumed to be constant from trigger time up to the nominal
source frame duration $T/(1+z)=20$ s of the gamma-ray emission,
and $t_i =\min[t/(1+z),T/(1+z)]$. We assume nominal values for the
initial temporal decay index $b1 \sim -1.1$, and for the index
after the jet break $b2 \approx -2$. For the energy spectral index
we adopt a nominal value $a \approx -0.7$, and $D_{l}$ is the
luminosity distance, using $\Omega_{tot}=1,\Omega_{m}=0.3, \rm{and}\
\Omega_{\Lambda}=0.7$.

Models for the time-dependence of the equivalent width ($EW$)
of the X-ray lines involve a number of physical and geometrical
assumptions (Ghisellini, Lazzati \& Campana 1999; Lazzati,
Campana, \& Ghisellini 1999; Weth et al 2000; \Mesz \& Rees
2000; Lazzati et al 2002; Rees \& \Mesz 2000; Kallman, \Mesz \&
Rees 2003). We do not intend here to delve
into the details of these models, setting ourselves instead a
simpler goal. That is, assuming that the lines so far detected are
real and representative, we ask ourselves up to what redshifts
and at what times would such lines be detectable with X-ray
instruments available in the next few years (in particular, taking
advantage of Swift's fast response time).
In what follows we adopt a phenomenological description of the
equivalent width of Fe group lines in the afterglow, based on the
reported detections in five cases of emission lines with equivalent
widths $EW' \sim 0.5-1$ keV at observer epochs $t\sim 0.5-1$ day
after the trigger. Earlier measurements with a large area high
resolution instrument do not exist (although one prompt absorption line
lasting of order $10$ s was reported with the wide-field detector
on Bepoo-SAX, Amati, etal, 2000), and this could be due to slewing
time limitations of previous missions. However, one expects in a
distant reprocessor (e.g. supranova) model the line to become
prominent at 0.5-1 day due to the geometry of the model (Lazzati
et al 1999; Weth et al 2000). On the other hand, in nearby
reprocessor (e.g. stellar funnel) models, a crude argument
indicates that emission lines could start as early as minutes after
the trigger (\Mesz \& Rees 2001), and the $EW$ may remain roughly
constant for times of order  one day, due to a long-lived decaying
jet or ourflow. Ballantyne \& Ramirez-Ruiz (2001) calculated in
more detail the evolution of $EW$ with incident luminosity, using
the reflection code developed by Ross, Weaver, \& McCray (1978)
and updated by Ross \& Fabian (1993). Using solar abundances and
incidence angles of 45 and 75 degrees, they find for both distant
and nearby models a similar $EW$ tendency of an initial increase
as a power law until reaching a plateau maximum, followed by a
steeper decay. For the distant model, this can be understood as
being due to the ionization parameter $\xi=L_x/n r^2$ being in the
range $10\time 10^2-3\times 10^3$ where high-ionization Fe lines
are prominent (for solar abundances) at the time when the effective
emitting area dictated by the light travel time to the shell
reaches a maximum, and afterwards dropping off. For the nearby
model, the similar behavior in this calculation may be ascribed to
the $L_x$ drop in time, $\xi$ initially being too high for Fe lines,
then after a decrease being for a finite time into the optimum range
for high-ionization Fe lines, and then dropping below the optimum range.
Thus qualitatively this calculation suggests that both models have
a rising and decaying $EW$, with a peak near one day, compatible
with current observations.

There are a  number of factors, however, which can lead to
significant changes in the simple model discussed above.  For
instance, more shallow incidence angles, as might be expected in
nearby funnel models, can increase significantly the $EW$ over
what is obtained at wider incidence angles (Kallman, \Mesz \&
Rees, 2003), and the $EW$ also increases when the Fe abundance
is larger than solar. A supersolar Fe abundance also
affects the luminosity dependence of the ionization equilibrium;
as shown by Lazzati, Ramirez-Ruiz \& Rees (2002), an Fe abundance
ten time solar extends to $3\times 10^1 \lesssim \xi\lesssim 10^5$
the range where highly ionized Fe lines are  prominent. Thus, for
a typical nearby funnel model with $\sim 1 M_\odot$ inside a shell
$\Delta r/r=10^{-1}$ at $r\sim 10^{13}r_{13}$ cm and density
$n\sim 10^{18}n_{18}$ cm$^{-3}$, an initial luminosity $L_x\sim
10^{50}$ erg/s at source time $T'=10$ s decaying $\propto
t^{-1.2}$ leads to a prominent Fe line (for ten times supersolar
Fe) for $10^5 \gtrsim \xi \gtrsim 30$ between source times $10^2
\lesssim t' \lesssim 10^5$ s, corresponding at $z=9$  to observer
times $10^3 {\rm s}\lesssim t \lesssim 10^6 \ {\rm s}$. For 100
times supersolar Fe, the Fe lines are expected to be prominent
starting at even shorter times. A funnel model can be expected to
be metal enriched and to have very shallow incidence angles. Thus,
based on the above simple argument, it is plausible to assume that
in a funnel model a large $EW$, say of order the 0.5-1 keV values
reported, could be present starting minutes after the trigger and
up to days. Such approximate estimates, of course, would need more
careful testing via numerical calculations, requiring a number of
additional assumptions and extensive parameter space modeling.
Short of doing that, we can bracket the line $EW$ behavior of funnel
models as ranging between, at one extreme, being similar to that of
the distant models (that is, an $EW\sim 1$ keV only near 0.5-2 days),
and at he other extreme, having an almost constant $EW \sim 1$ keV
from minutes to days. Clearly, a
distinction between nearby or distant models will require much
further numerical modelling, and/or observations determining
whether a dense reprocessor can be present at $r\gtrsim 10^{16}$
cm at the time of the burst. We do cannot address such a choice
here. However, noting that the generic behavior of a line
producing region is bound to be bracketed between the above
mentioned two extremes, we investigate the line detectability in
the case of both of these behaviors.

The simplified approach adopted here is to assume a phenomenological
X-ray continuum whose time behavior is given by equation (\ref{eq:Fxdouble}),
and assume that the afterglow produces Fe group lines whose rest-frame
$EW'\sim 1$ keV is comparable to the reported values, without specifying
the physical model giving rise to them. For the line temporal behavior we
consider the two cases above. One case has an (approximately) constant
$EW$ between minutes to days, which may be plausible for nearby (funnel)
models under conditions as discussed above. The other case treated
assumes a variable $EW$, which starts small and peaks around one day
(based on the calculations of Ballantyne \& Ramirez-Ruiz, 2001, e.g.
their intermediate curve,
where $EW$= 10 eV  for $L_x>10^{48}$ erg/s; then an $EW$ increasing as a
power law for $10^{46.5}>L_x> 10^{48}$ erg/s, up to  $EW \sim 1$ keV for
$10^{45.5}<L_x<10^{46.5}$ erg/s; and an $EW$ exponential decay for
$L_x< 10^{45.5}$ erg/s. This implies an $EW$ peak time around
$\sim$1-2 days at redshift $z\sim 1$ in our model. For these two models,
we then calculate the signal to noise ratio of the line observation as a
function of redshift and observer time.

\begin{figure}[ht]
\centerline{\psfig{file=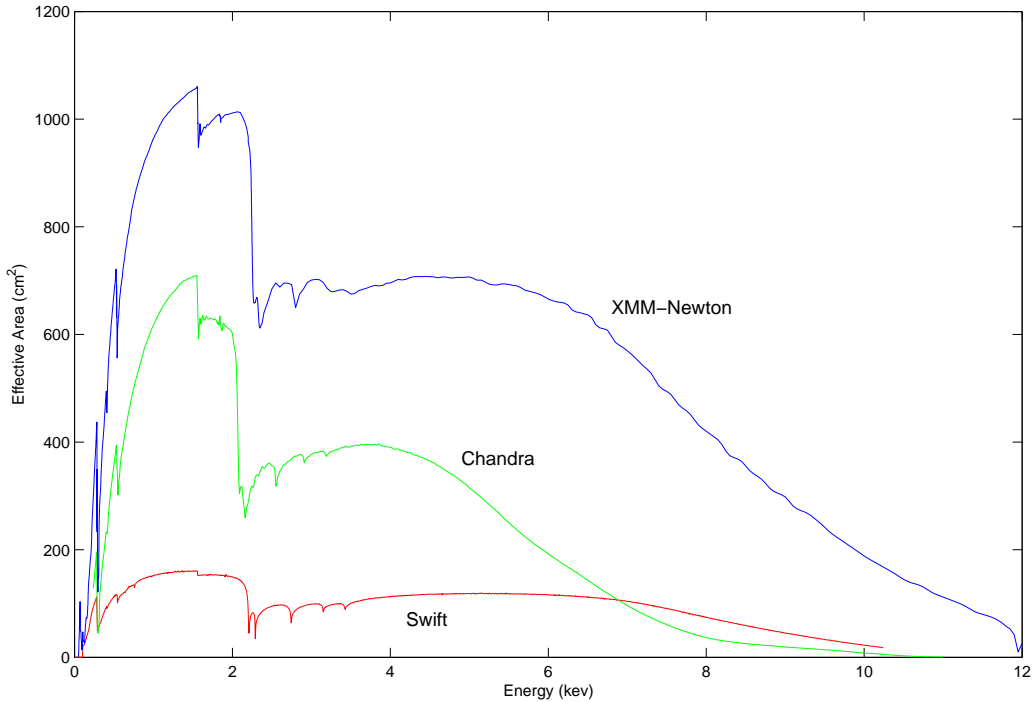,scale=0.6,angle=-90}}
\caption{Effective area of the Swift XRT, Chandra-ACIS and
XMM-Epic detectors} \label{fig:effarea}
\end{figure}

 The typical procedure which we follow is: (i) we create a
nominal observed spectrum from the theoretical equation
(\ref{eq:Fxdouble}) and an emission line of a given $EW$ and assumed
width (e.g. due to thermal motions or bulk dispersion velocities),
and convolve this with the response function of the instrument,
using the standard X-ray spectral fitting package {\it XSPEC}.
As examples, we have used the response functions of Swift XRT,
Chandra ACIS and XMM-epic. The relative effective areas of these
instruments are shown in figure \ref{fig:effarea}.

In practice, the input spectrum is a power-law continuum plus
a gaussian $K \alpha$ iron line, taking into account the absorption
by the galactic medium (wabs*(powerlaw+gaussian)). For this model,
we have to input 6 initial parameters: (a) galactic column
density, (b)power-law continuum power-law index, (c)continuum flux
normalization factor, (d) gaussian line energy, (e) line width,
and (f) normalization for the gaussian line. The galactic column
density was set to a typical value $2\times 10^{20}\ {\rm
cm^{-2}}$ (N. Brandt 2003, private communication). The continuum
normalization is determined by the continuum flux given by the
formula (\ref{eq:Fxdouble}), making those two fluxes consistent
over the instrument observing band (e.g. 0.2-10 keV). Because the
simulated spectrum is seen in the observer frame, the Gaussian
line energy is given by the equation $E_{l}=6.7\ {\rm keV}/(1+z)$.
The line width is taken to be proportional to the line energy.
Observationally, there is no consensus on
how much the ``typical" line width is. For the same burst GRB
991216, Piro (2000) obtained a line width of $\sigma > 1$ keV,
while Ballantyne et al. (2002) used a narrow line ($\sigma=0.01$
keV) to get a better fit. However, it is reasonable to assume it
to be a broad line, because physically the photoionization of
Fe-rich plasmas for an ionization parameter $\xi \sim 10^{3}$ is
expected to result in an equilibrium rest-frame temperature of a
few keV. In the simulations, we have used a fixed relation between
the line energy and the line width, which for the most part is
$\sigma_{l}=0.1\ E_{line}$ keV (except where we compared broad vs.
narrow lines).The last parameter (f) is the equivalent width $EW$,
which for the constant $EW$ models is set to be 1 keV,
and in the varying $EW$ case it is given by the time dependence
discussed above leading to a value $\sim 1$ keV at about one day.

(ii) Once we have the simulated spectrum, we follow the standard
method of analyzing the spectrum using the {\it XPSEC} software
package and determine to what degree the line is detectable,
calculating the significance level (in standard deviations
$\sigma$) for one specific detection according to the F-test.

(iii) We repeat the steps ((i) and (ii)) for at least 100 times,
and check the repeatability of the detection rate of a line at a
certain significance level.

In addition, we have also tested the degree to which instruments
can differentiate a broad line from a narrow line, or distinguish
between lines with different central energies, such as 6.4 and 6.7
keV. This is of interest as a diagnostic for the dynamics and
thermal conditions in the emission region, which would provide a
valuable tool in assessing models. The procedure used for these
two tests are similar to those described above. The simulated
spectra are obtained assuming the same input parameters as before.
Each simulated spectrum is fitted with a power law and a fixed
rest-frame line energy at either 6.4 keV or 6.7 keV, and we check
the chi-square difference between the two fits, which indicates
how the change of line energy affects the fit result. We repeated
this fit and comparison procedure for each set of parameters 50
times, in order to reduce the statistical errors,
and derive from this an average chi-square difference. Based on
the chi-square difference, we find the corresponding probability,
which is a function of the chi-square difference and the degrees
of freedom (Chapter 15.6, Numerical Recipes in C++, Press et al.
2002). For example, a chi-square difference of 21 and 5 degrees of
freedom corresponds to a probability 0.999, which is 3 $\sigma$.
For the narrow line versus broad line test, we fixed as examples
the line width to be 670 eV (broad line) and 200 eV (narrower
line), respectively, in the rest frame, and follow a similar
procedure as above for the line energy test.

Note that in our simulations we have used the chi-square
statistics throughout, instead of likelihood statistics, which is
valid and guaranteed by the large enough number of photons in each
bin (at least 20 photons) and in total (usually thousand of
photons collected during the whole integration time).

\section{Simulation Results: Line Detection Significance}
\label{sec:results}

We have done simulations for several sets of GRB afterglow model
parameters. The source-frame duration of the GRB is taken to be
$T'=20$ s or $T'=40$ s. We show only the plots for $T'=20$ s, the
longer durations  being used only for comparison. The initial
isotropic-equivalent luminosity is usually taken to be
$L_{X,0}=10^{50} \ {\rm ergs\ s^{-1}}$, consistent with the
present ($z\siml$ few) observations (Costa 1998). As an
alternative, we also consider a higher than usual initial X-ray
luminosity, $L_{X,0}=10^{51}$ erg s$^{-1}$, which might be typical
of high redshift GRB, e.g. Schneider et al, 2002). (Note that we
take an initial X-ray luminosity which is assumed to be about one
order of magnitude below the corresponding prompt $\gamma$-ray
luminosity). The source-frame equivalent width is typically taken
as $EW'=1.0$ or 0.5 keV (only the 1 keV results are shown).
We consider the two generic line temporal behaviors discussed
above, one in which the K$\alpha$ line has a a large $EW'\sim 1
{\rm keV} \sim$ constant from minutes after the trigger up to days,
and another in which the $EW'$ starts low, grows to $EW'\sim 1$  keV
on a timescale $\sim 1$ day, and drops rapidly afterwards
(see discussion in \S \ref{sec:model}).
The integration time was taken to be 0.6 times the observing time,
counted after the GRB trigger (i.e. for an observer time 1 day
we take we take an integration time of 0.6 days ending at 1 day.

We took a grid of values in redshift $z$ and observer time $t$, and
with the above parameters and the procedure outlined in the previous
section we calculate for each pair of $z,t$ values the significance
level of the detection with various instruments. For most of the
calculations (unless stated otherwise) this procedure is repeated 300
times for each point, and the average chi-squared value is adopted,
resulting in contour plots of the standard deviation $\sigma$ in the
$z,t$ plane. E.g., regions in the plot with significance levels
$\simg 3\sigma$ indicate that the detection is likely to be real.

\begin{figure}[ht]
\centerline{\epsfig{file=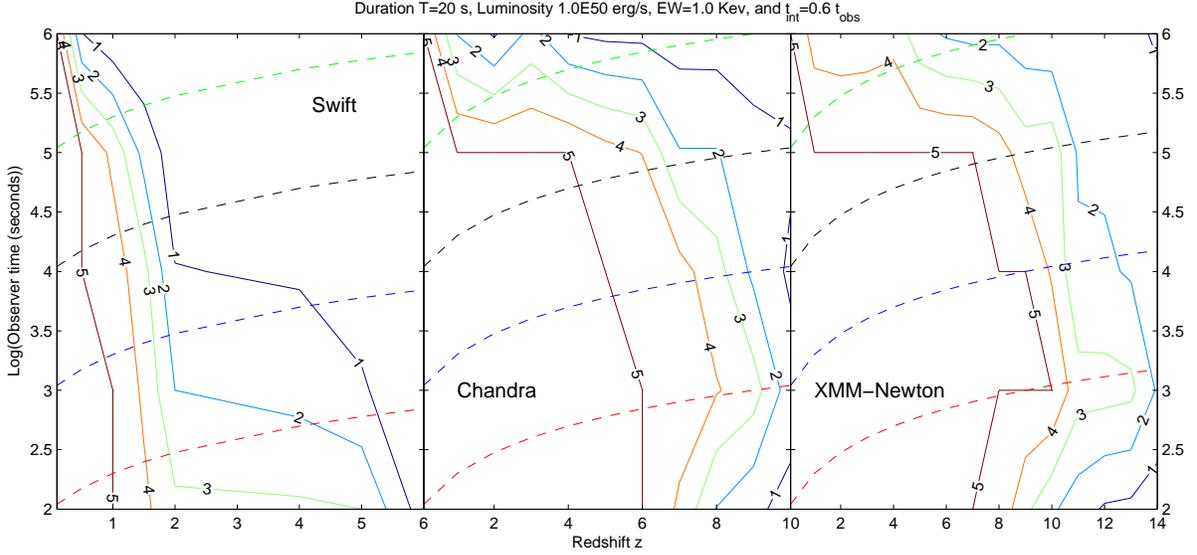,angle=-90, scale=0.6 }}
\caption{Fe K-$\alpha$ detection significance level contour plot
for Swift (left), Chandra (middle) and XMM (right), assuming a
source-frame GRB duration $T'=20$ s and a constant $EW'=1.0$ keV,
initial X-ray luminosity $L_{X,0}=10^{50}\ {\rm ergs}\ {\rm
s^{-1}}$ and integration time $t_{int}=0.6\ t_{obs}$. The dashed
lines indicate contours of constant source-frame time.}
\label{fig:swchxm_50_1.0}
\end{figure}

Fig \ref{fig:swchxm_50_1.0} shows the line detection significance
levels attainable with Swift, Chandra and XMM for a burst of
initial X-ray luminosity $L_{X,0}=10^{50}\ {\rm ergs}\ {\rm
s^{-1}}$, source-frame equivalent width $EW'=1.0$ keV (assumed
constant), for a source-frame GRB prompt duration $T'=20$ s. This
shows that Swift can detect such iron lines with significance
$\simg 4\sigma$ up to $z \siml 1.5$ for observer times $t\siml
10^3$ s, or up to $z \siml 1.2$ for observer times $t\siml 10^4$
s. to $z\siml 1$ for observer times up to a day. The bends in
the significance level plots, e.g. at intermediate redshifts and
times for Swift, are due to the minima in the effective area of
the instrument at intermediate energies (figure
\ref{fig:effarea}). This is superposed on the expected overall
tendency of a decreasing significance level with increasing time
and redshift. For longer burst durations, e.g. $T'$=40 s, the
continuum flux level is correspondingly higher at the same time,
and the lines are detectable to correspondingly higher redshifts
or times. For a decreased equivalent width, e.g. $EW'=$0.5 keV,
Swift can detect such iron lines only in very nearby ($z \lesssim
0.3$) afterglows at a significance level of $\simg 4\sigma$, at
observer times up to a day.

\begin{figure}[ht]
\centerline{\epsfig{file=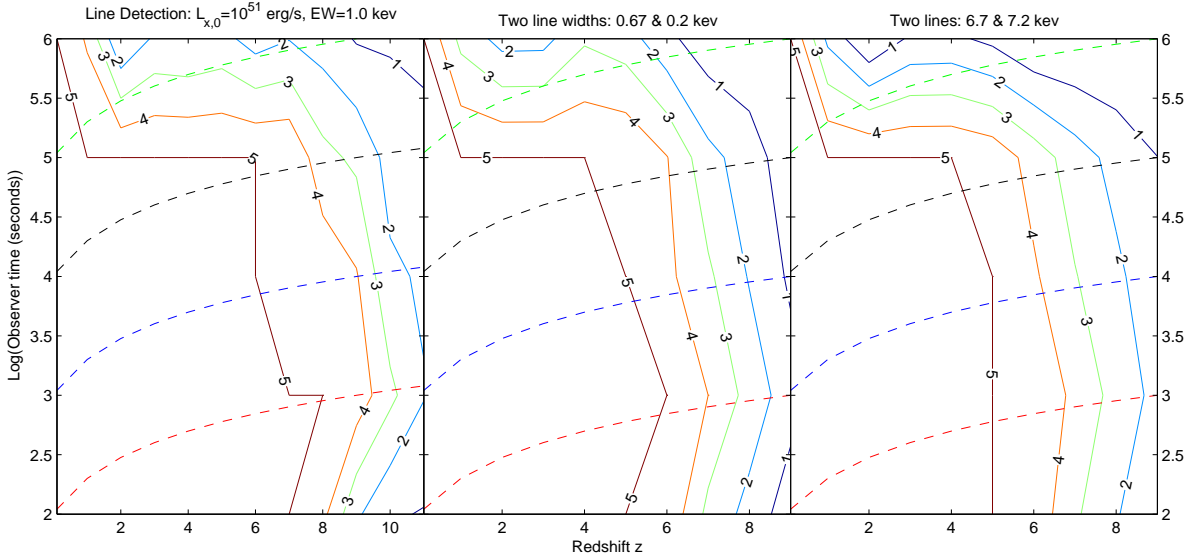,angle=-90,scale=0.6}}
\caption{Bursts of higher initial luminosity $L_{X,0}=10^{51}$
ergs s$^{-1}$ observed by Swift. Shown is the ability to detect an
Fe-$\alpha$ line (Left panel). Also shown is the ability to
distinguish Fe 6.7 keV lines of different widths $\Delta E$=0.67
keV vs. $\Delta E$=0.2 keV (Middle panel), and to distinguish two
lines of different central energies 6.7 keV (He-like Fe) vs.  7.2
keV (He-like Co) (Right panel). The contour levels give the
significance for separating the two lines. In all cases, $EW'=1.0$
keV, GRB duration $T'=20$ s, and integration time $t_{int}= 0.6\
t_{obs}$.} \label{fig:merge_all_sw_51_1.0}
\end{figure}

The same calculations for the same burst with the Chandra ACIS
detector (Fig.  \ref{fig:swchxm_50_1.0}, middle) and the XMM Epic
detector (Fig \ref{fig:swchxm_50_1.0}, right) show a significantly
greater depth of detection, as is to be expected from the larger
effective areas. In all cases, the significance levels become
lower as the observing time increases, as expected from the
dimming, hence an earlier acquisition of the target as well as an
early turn-on of the line improve the chances of detection. These
plots also show features which are due to the detector
characteristics. Roughly, one can say that bursts with the
standard parameters used here would be detectable at better than
$4\sigma$ confidence level with Chandra up to $z\siml 6-6.5$, and
with XMM up to $z\siml 8.5-9$, at observer times $\siml 10^5$ s = 1
day. For later observer times, similar significance may be
obtained for lower redshifts.

Fig \ref{fig:merge_all_sw_51_1.0} (left panel) shows the Swift
line detection ability for a higher initial X-ray luminosity case,
$L_{X,0}\sim 10^{51}\ {\rm erg\ s^{-1}}$, and constant $EW'$ =1.0 keV.
Such values may occur occasionally at moderate to low redshifts, and
may also characterize bursts from very massive pop. III stars at
large redshifts $z\simg 6$ (Schneider et al.  2002). For such higher
initial fluxes, Swift may be able to detect lines at those high
redshifts, while Chandra and XMM would do even better. Results for
Swift in Fig. \ref{fig:merge_all_sw_51_1.0} (left panel) show that
Fe lines would be detectable to better than $4\sigma$ up to $z\siml 8$
for observed times $t\siml 10^5$ s. (For $EW'=0.5$ keV, not shown, this
significance is achievable only to $z\siml 3.5$ and $t\siml 10^4$ s).
There is a significant difference between the redshifts for, say, a
$5\sigma$ line detection in the $L_{x,0}=10^{50}$ erg/s and for
$L_{x,0}=10^{51}$ erg/s cases. For the lower luminosity case, this
significance extends only up to $z\sim 1$, while for the higher
luminosity the same significance level is reached up to $z\sim 7$.
The difference is due mainly to two factors. First, the K-correction
factor $(1+z)^{-1-a+b1} \sim (1+z)^{-1.4}$ (for the parameters used
in the paper) compensates in part for the flux reduction as the
luminosity distance increases with redshift. Second, there is a
detector effective area difference between low and high redshift
lines, which from Fig. \ref{fig:effarea} is seen to a factor $\sim 2$
increase in effective area for lines at $z\sim 7$ relative to
those at $z\sim 1$.

\begin{figure}[ht]
\centerline{\psfig{file=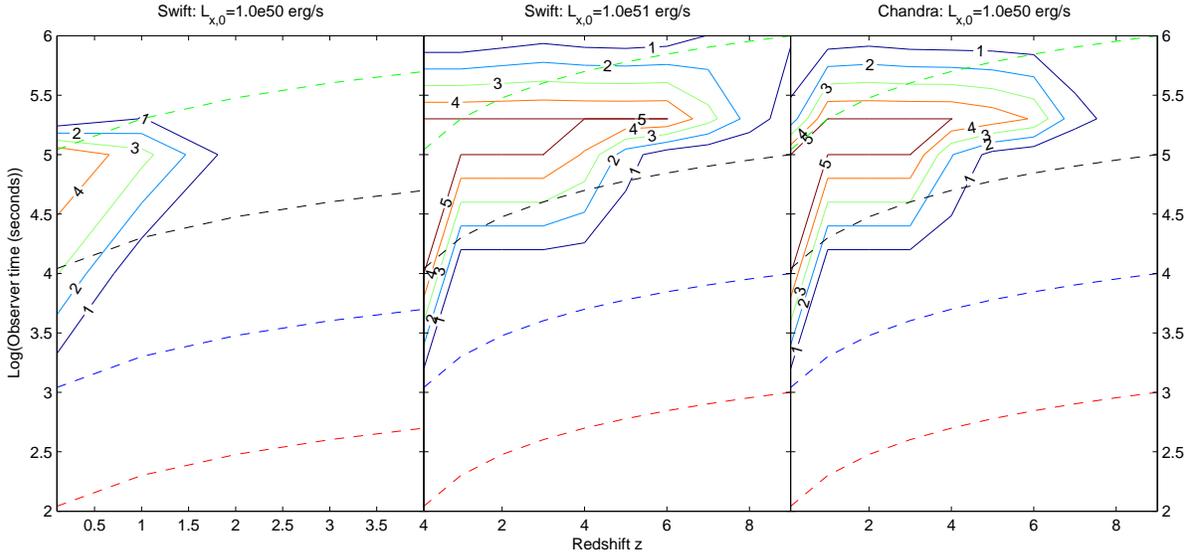,angle=-90, scale=0.6 }}
\caption{Variable equivalent width case, for a standard luminosity
burst $L_{X,0}=10^{50}$ erg/s seen with Swift (left), a higher
luminosity burst $L_{X,0}=10^{51}$ erg/s seen with Swift (middle),
and a standard luminosity burst $L_{X,0}=10^{50}$ erg/s seen with
Chandra (right). Shown are the line detection significance
contours, in all cases the $EW$ peaks at 1 keV at one day in the
rest frame (see text). Other parameters and symbols are the same
as in Fig. \ref{fig:swchxm_50_1.0}}. \label{fig:ew_vary}
\end{figure}

Fig \ref{fig:ew_vary} shows the line detectability by Swift
in the case of a varying $EW$, using the $EW$ behavior with time
calculated by Ballantyne \& Ramirez-Ruiz, as discussed in \S
\ref{sec:model}. The most prominent characteristic on this plot is
that the region where the detection confidence level is high lies,
as expected, inside a strip corresponding to source times between
$10^{4.5}$ and $10^{5.5}$ s. For the case of Swift and the standard
initial luminosity $L_{X,0}=10^{50}\ {\rm ergs}\ {\rm s^{-1}}$ the
$4\sigma$ level is achieved for $z\siml 0.7$ and observer times
$t\sim 10^{4.5}-10^5$ s, while for the higher $L_{X,0}=10^{51}\
{\rm ergs}\ {\rm s^{-1}}$ case this is achieved at $z\siml 6$ and
$10^{4.5}-10^{5.5}$ s (left and middle panels of Fig.
\ref{fig:ew_vary}). Most of the area on the plot outside this
ridge shows a low significance, as expected, since in this case
for initial observer time $\sim 10^4$ s, and also after $\sim
10^{5.5}$ s, the iron lines have a low $EW$ and are too weak to be
detected. Comparing with the $EW$-constant case for the same
parameters, the detectability of the varying $EW$ result near the
peak is consistent with what is obtained in the constant $EW$ case.
With Chandra, the same varying $EW$ case but with the standard
initial luminosity $L_{X,0}=10^{50}\ {\rm ergs}\ {\rm s^{-1}}$
indicates that detection at the $4\sigma$ level can be achieved up
to $z\siml 5.5$ at $t\sim 10^{4.5}-10^{5.5} $ s.

\begin{figure}[ht]
\centerline{\psfig{file=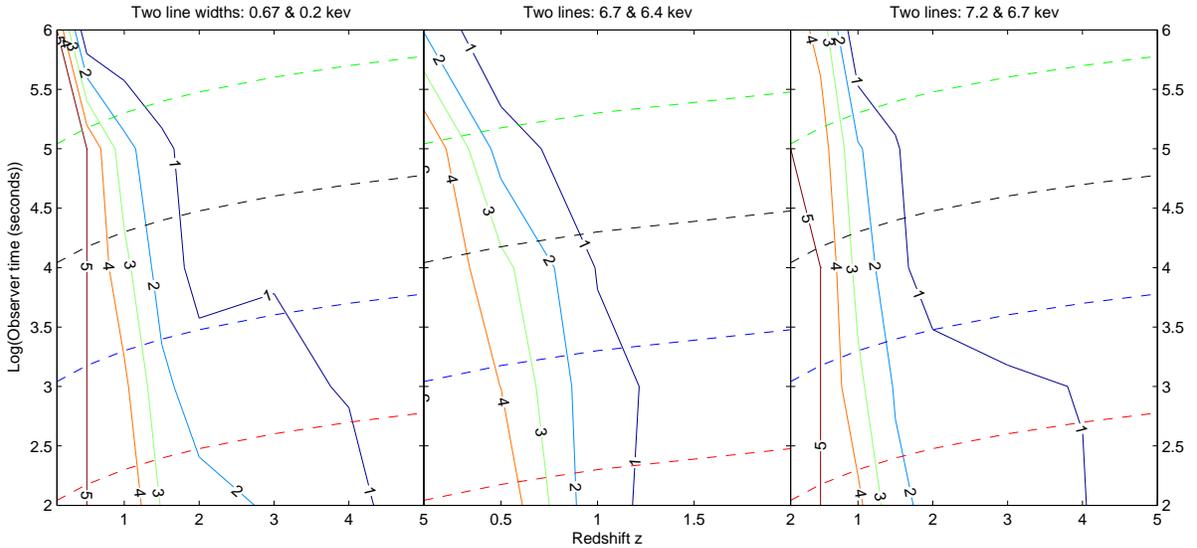,angle=-90, scale=0.6}}
\caption{Swift ability to distinguish two Fe 6.7 keV lines of
widths 0.67 keV vs. 0.2 keV (Left), or two Fe lines of the same
width 0.67 keV and central energies 6.4 keV vs. 6.7 keV (Middle),
or two lines of the same width but central energies 6.7 keV
(He-like Fe) vs.  7.2 keV (He-like Co) (Right). The contour levels
give the significance for separating the two lines, for standard
bursts of initial X-ray luminosity $L_{X,0}=10^{50}$ ergs
s$^{-1}$, $EW'=1.0$ keV, GRB duration $T'=20$ s, and integration
time $t_{int}= 0.6\ t_{obs}$.} \label{fig:merge_all_sw_50_1.0}
\end{figure}

Another useful calculation is to find the maximum redshifts
and times for which various instruments can still distinguish a
narrow from a broad line, or between different line energies, say
between Fe 6.4 keV vs. 6.7 keV, or between corresponding Fe and Co
K-$\alpha$ lines. For determining the significance level to which line
differences can be measured, we follow the method discussed in \S
\ref{sec:model}. In principle, one might expect that when the line
broadening decreases below the nominal energy resolution of a
detector, one cannot differentiate between widths below this value.
The resolution of the Chandra ACIS S3 instrument is $\sim 0.1$ keV,
and that of the Swift XRT detector is $\sim 0.3$ keV. Thus with
Chandra, for a line width of 0.67 keV, or 10\% of the rest frame
line energy, the observed line width for a GRB at $z\sim 6$ would
be at the nominal energy resolution limit. However, even if this
redshift is exceeded, or if the line is narrower, different width
lines may still be distinguishable in a statistical way.
The distinguishability of different line broadenings or line
energies is affected by several additonal factors, such as the
$EW$, the integration time, the degree to which the centroid of
the lines can be characterized, etc.

\begin{figure}[ht]
\centerline{\psfig{file=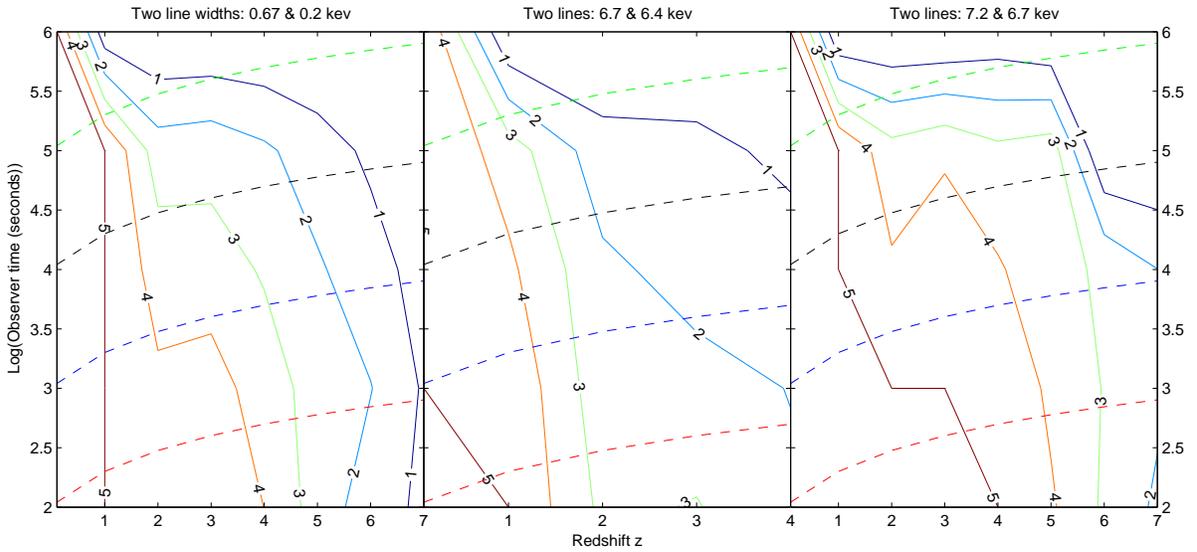,angle=-90, scale=0.6}}
\caption{Chandra ability to distinguish two Fe 6.7 keV lines of
widths 0.67 keV vs. 0.2 keV (Left), or two Fe lines of the same
width 0.67 keV and central energies 6.4 keV vs. 6.7 keV (Middle),
or two lines of the same width but central energies 6.7 keV
(He-like Fe) vs.  7.2 keV (He-like Co) (Right) . The contour
levels give the significance for separating the two lines, for
standard bursts of initial X-ray luminosity $L_{X,0}=10^{50}$ ergs
s$^{-1}$, $EW'=1.0$ keV, GRB duration $T'=20$ s, and integration
time $t_{int}= 0.6\ t_{obs}$.} \label{fig:merge_all_ch_50_1.0}
\end{figure}

\begin{figure}[ht]
\centerline{\psfig{file=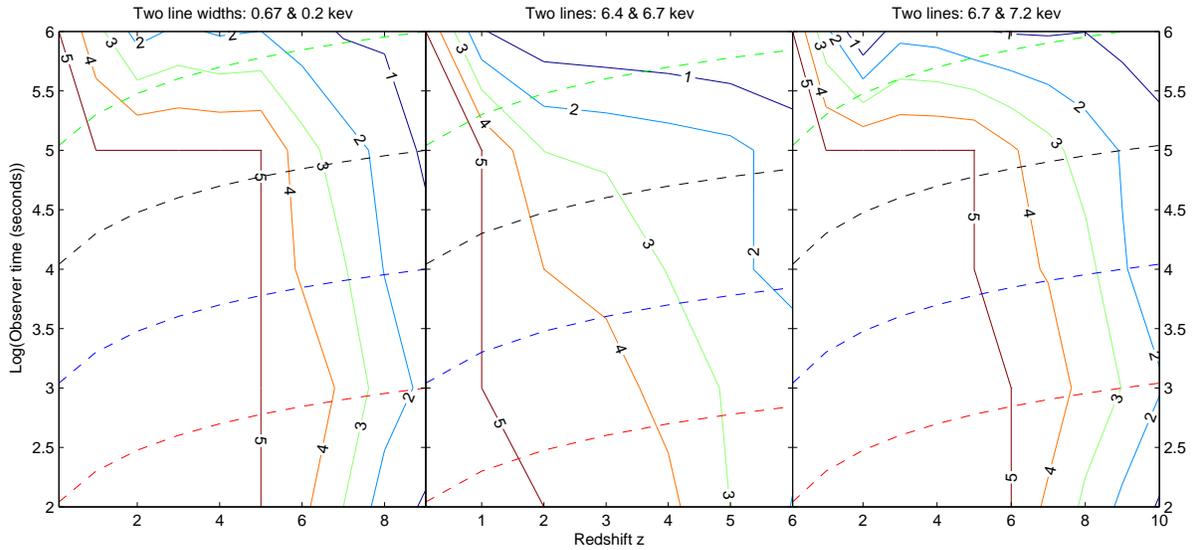,angle=-90, scale=0.6}}
\caption{XMM-Newton ability to distinguish two Fe 6.7 keV lines of
widths 0.67 keV vs. 0.2 keV (Left), or two Fe lines of the same
width 0.67 keV and central energies 6.4 keV vs. 6.7 keV (Middle),
or two lines of the same width but central energies 6.7 keV
(He-like Fe) vs.  7.2 keV (He-like Co) (Right). The contour levels
give the significance for separating the two lines, for standard
bursts of initial X-ray luminosity $L_{X,0}=10^{50}$ ergs
s$^{-1}$, $EW'=1.0$ keV, GRB duration $T'=20$ s, and integration
time $t_{int}= 0.6\ t_{obs}$.} \label{fig:merge_all_xmm_50_1.0}
\end{figure}

In practice, using {\it XSPEC} and the averaging of multiple tries
described above, the maximum redshift to which lines of different
broadenings can be differentiated is found to be somewhat larger
than what is expected from the simple estimate above. In our
simulations, we have taken a nominal ``broad" line width of 0.67
keV (10\% of the line energy) and a nominal "narrow" line width of
0.2 keV, for the Fe K$\alpha$ 6.7 keV line. The results for Swift
are shown in the left panel of Fig \ref{fig:merge_all_sw_50_1.0},
while the results for Chandra are shown in the left panel of Fig
\ref{fig:merge_all_ch_50_1.0} and those for XMM are shown in the left
panel of Fig \ref{fig:merge_all_xmm_50_1.0}, for the standard
luminosity burst case.

For lines of different central energies, such as He-like Fe
6.7 keV K$\alpha$ and the lower ionization state Fe 6.4 keV line,
the line energy difference is 0.3/(1+z) keV, and naively one
expects a sensitivity to detecting different lines up to a
redshift $z\sim 2$. The same factors, such as $EW$, integration
time, etc, will affect the detectability. Using the statistical
method described above, we have plotted for Swift in
Fig \ref{fig:merge_all_sw_50_1.0}, for Chandra in
Fig \ref{fig:merge_all_ch_50_1.0}, and for XMM-Newton in
Fig \ref{fig:merge_all_xmm_50_1.0} the significance contours for
differentiating a 6.4 vs. a 6.7 keV line (middle panels).

We have also investigated the ability to distinguish
between the He-like Fe 6.7 keV and the corresponding Co 7.2 keV or
Ni 7.8 keV lines (which are all He-like) . We find that it is
easier to distinguish Co or Ni He-like lines from the Fe 6.7 keV
line than it is to distinguish between the Fe 6.7 and 6.4 keV
lines, because the line energy differences are larger than that
between Fe 6.4 and 6.7 keV lines. The significance contours for
differentiating an Fe 6.7 keV from a Co 7.2 keV line are shown for
Swift in Fig \ref{fig:merge_all_sw_50_1.0}, for Chandra in
Fig \ref{fig:merge_all_ch_50_1.0}, and for XMM-Newton in
Fig \ref{fig:merge_all_xmm_50_1.0} (right panels, in all three),
for a standard burst of initial X-ray luminosity $L_x=10^{50}$ erg/s,
$EW'=1.0$ keV, $T'=20$ s. For bursts of higher initial luminosity
$L_{X,0}=10^{51}$ erg/s, Swift's ability to discriminate between
Fe 6.7 keV and Co 7.2 keV lines is shown in the right panel of
Fig \ref{fig:merge_all_sw_51_1.0}.

We note that if lines of different widths or different energies
are determined to be distinguishable with the above procedure,
this implies that the lines are detectable at the 3 $\sigma$ or
higher level. This is because in {\it XSPEC}, if a line is not
detectable or is detectable at a smaller significance level, the
possible parameter range does not make a significant difference in
the fitting results. The converse, however, is not true, since
there may be cases where a line is detectable at the 3 $\sigma$
level, while differences in the line energies or line widths
cannot be distinguished at a comparable level of significance.

\section{Discussion}
\label{sec:disc}

We have quantitatively investigated the prospects for the
detection of Fe group X-ray emission lines in gamma-ray burst
afterglows, for various redshifts extending up to $z\sim 14$ and
observing times extending up to 10 days after the trigger.
We have used as a template for the X-ray continuum the spectral
and temporal behavior inferred from afterglow observations, and
have assumed Fe line rest-frame equivalent widths $EW'\sim 1$ keV
comparable to those reported in a number of GRB afterglows.
These line and continuum models are purely phenomenological,
without explicit reference to particular models. It is not known
at present whether lines, if present, are formed soon after the
outburst or only after periods of a day or so, as suggested by
existing observations, since slewing time constraints in previous
experiments prevented verification of the time of onset of the lines.
Thus, we have tested for the line detectability at observer times
starting at $t\sim 100$ s and up to days. We have assumed two
simplified models for the equivalent width time behavior, motivated
by generic analytical and numerical photoionization models. One of
these assumes an equivalent width which does not change significantly
with time, being a constant fraction of the photoionizing continuum.
This model may be pertinent to so-called nearby models of line
formation in the jet funnel or outer parts of the expanding star.
The other assumes an equivalent width which grows in time up to a
maximum value reached at about one day, and subsequently declines.
This is motivated by so-called distant, or geometric models, e.g.
where a shell of dense material is encountered at about a light day
from the source. A similar behavior, however, may occur for some
values of the parameters also in nearby  models.

The results for simulated observations with the X-ray detectors in
Swift (XRT), Chandra (ACIS S3) and XMM (Epic) are presented as
contour plots of the significance level of line detection,
expressed in standard deviations, for a Gaussian line of constant
or varying equivalent width superposed on a power law continuum
declining in time, as a function of redshift and observing time.
As expected, the line detectability is sensitively dependent on
the initial and subsequent continuum X-ray luminosity and on the
equivalent width. Due to the K-correction effects whereby the spectral
and temporal redshift-dependent changes partially cancel out, the
sensitivity decrease in a given energy band with increasing redshift
is more moderate than one would naively expect from the bolometric decay.
Notice that this is dependent on the afterglow spectral and temporal
indices entering in $F_E\propto E^a t^b$. For the nominal values
adopted here, $a=-0.7$, $b=b1=-1.1$ (before the break), the K-correction
is $(1+z)^{-1.4}$, but the specific values can differ substantially
between bursts.

For an initial X-ray isotropic-equivalent luminosity
$L_{X,0}=10^{50}$ erg/s, a constant source-frame equivalent width
$EW'=1.0$ keV and a prompt-phase source-frame duration $T'=20$ s,
Swift, Chandra, and XMM can detect lines  with significance $\simg
4\sigma$ roughly out to $z\siml 1.5,\ 6, {\rm and}\ 8.5$
respectively (Fig \ref{fig:swchxm_50_1.0}), for times of order
$t\siml 10^3,~10^{5},~10^{5}$ s. For a similar initial luminosity
but an equivalent width $EW'=0.5$ keV, the corresponding redshifts
drop substantially; e.g. Swift and Chandra could detect lines to
$z\siml 0.3\  {\rm and} \ 2$, respectively, with the same
confidence level as above for times of $t\siml 10^3,~10^5$ s.

In a more detailed model, the equivalent width could vary in
time, and in this case one would expect the line detectability to
be maximal when the equivalent width reaches its peak value (See
\S \ref{sec:model} for detailed model). Outside the peak range,
the $EW$ is smaller and the detectablity drops rapidly. Our
calculations for a standard luminosity burst of $L_{X,0}=10^{50}$
erg/s indicate that Swift could detect Fe lines up to $z \siml
0.7$ for times $t\siml 10^{4.5}-10^5$ seconds (Fig
\ref{fig:ew_vary}, left). For the same parameters, Chandra could
detect Fe lines out to $z \siml 5.5$ for times of $\siml
10^{4.5}-10^5$ seconds (Fig \ref{fig:ew_vary}, right). For more
luminous afterglows of $L_{X,0}=10^{51}$ erg/s (which may be
characteristic of pop. III bursts), Fe lines could be detected
with Swift out to $z \siml 6$ with the same confidence level and
observer time range as Chandra does for the standard (lower)
luminosity case.

An interesting question is how far and how late can various
instruments distinguish between lines of the same energy centroid
but different widths, or between lines of similar widths but
different energy centroids. The former is more difficult,
especially at high redshifts or late times when the detection is
marginal, and for this we investigated the distinguishability of
nominal "broad" lines with $\Delta E/E\sim $ 0.1 vs. "narrow"
lines with $\Delta E/E\sim $ 0.03, i.e. 0.6 keV vs. 0.2 keV
widths for the Fe 6.7 keV lines. For investigating
the sensibility to different line centroids, we assumed the same
widths $\Delta E=0.67$ keV for lines of centroid energies 6.7
keV vs. 6.4 keV (He-like Fe vs. the lower ionization Fe line
complex), as well as 6.7 keV (He-like Fe) vs. 7.2 keV (He-like
Co). For a standard luminosity afterglow of $L_{X,0}=10^{50}$
erg/s and a constant equivalent width $EW'=1.0$ keV, $T'=20$ s,
the figures (\ref{fig:merge_all_sw_50_1.0}),
(\ref{fig:merge_all_ch_50_1.0}) and
(\ref{fig:merge_all_xmm_50_1.0}) show the ability to distinguish
these different line cases with Swift, Chandra and XMM-Newton,
respectively. One sees that for such standard luminosity bursts,
Swift can differentiate an 0.67 keV  line width from an 0.2 keV
width out to $z\siml 1.2$ with confidence level of $\siml 4\sigma$
for times $t\siml 10^{3.5}$ s. It it can distinguish a 6.7 keV
line from a 6.4 keV line out to $z \siml 0.2$ with confidence
level $\sim 4\sigma$ at times $t\siml 10^5$ seconds;  and a 6.7
keV line from a 7.2 keV line out to $z\sim 0.75$ with $\siml
4\sigma$ for times $t\siml 10^5$ s. As might be expected, a 7.2
keV line is easier to distinguish from a 6.7 keV than the latter
is from a 6.4 keV line, since the energy difference is larger.
Because Chandra and XMM have much larger effective areas, one
expects that they can make such distinctions out to larger
redshifts and longer times, as is verified from an inspection of
fig \ref{fig:swchxm_50_1.0} (middle and right panels). E.g. with
Chandra, and $L_{X,0}=10^{50}$ erg/s, $EW'=1.0$ keV,
the same line width differences can be distinguished  out to
roughly $z\sim 1.5$ and $t\siml 10^{5}$ s with a confidence level
$\sim 4\sigma$ (Fig \ref{fig:merge_all_ch_50_1.0}, left), and
energy differences of 6.4 vs. 6.7 keV out to $z\sim 1$ and $t\siml
10^{4.5}$ s (Fig \ref{fig:merge_all_ch_50_1.0}, middle); while
energy difference of 6.7 vs. 7.2 keV can be distinguished out to
$z\sim 1.8$ and $t\siml 10^{4.5}$ s with $\sim 4\sigma$( Fig
\ref{fig:merge_all_ch_50_1.0}, right). For XMM, the corresponding
redshifts are of order is $z \siml 5.5, \ 1.5, \ 6$ with a
confidence level of $\siml 4\sigma$ for times of order $10^{5}$ s
(Fig \ref{fig:merge_all_xmm_50_1.0}).

For larger initial isotropic-equivalent X-ray luminosities,
e.g. $L_{X,0}=10^{51}$ erg/s corresponding to  an extreme
low-resdhift case or a nominal high-redshift pop. III case, Swift
(and of course Chandra and XMM) can detect lines out to much
higher redshifts compared to the standard case of
$L_{X,0}=10^{50}$ erg/s.  This is shown in Fig
\ref{fig:merge_all_sw_51_1.0}, where for a constant $EW'=1.0$ keV
and $T'=20$ s  the maximum $4\sigma$ redshifts at $t\sim 10^5$ s
are $z \sim 8$, $6$, and $5.5$ for the detection of a 6.7 keV line,
for distinguishing two line widths of 0.67 vs. 0.2 keV, and for
distinguishing two lines of 6.7 vs. 7.2 keV (left, middle and
right panels, respectively).

In conclusion, if X-ray lines are present in GRB, Chandra and XMM,
with their slower slew response times, should be able to
detect them at observer times $\simg 0.5$ days ($0.5/(1+z)$ days
in the source frame) out to very high redshifts $z\sim 7-10$,
in the range where the universe started to reionize, for burst
properties similar to those inferred in $z\sim 1-3$ objects. If
XMM and Chandra do not detect new lines at a higher significance
level than previously reported (noting that line evidence is
considered currently for 9 out of 21 bursts, e.g. Sako et al.
2004), one might conclude that the conditions assumed here do not
apply to some or all of the bursts; or the $EW$ are smaller (e.g.
as in the variable $EW$ case, see Fig \ref{fig:ew_vary}); or the
luminosity is smaller; or the lines do not appear early on, when
the flux is high, etc. If on the other hand lines are detected,
XMM and Chandra should also be able to distinguish details such as
line widths or central energies out to redshifts where the first
galaxies formed, $z\sim 6$.  Swift, with its fast slew time,
should be able to detect lines at much earlier times, if present,
starting at $t\sim 10^2-10^3$ s, and out to redshifts $z\sim 1.5$,
or to higher redshifst for the more luminous bursts. In the latter
case, it would also be able to answer questions about line broadening,
ionization stage or line physics out to $z\siml 0.8$ and times up
to a day.

Thus Swift, in conjunction with larger spacecraft such as Chandra and XMM,
should be able to answer important questions about burst properties as
a function of the age of the universe, such as whether X-ray lines occur
in them, at what times in the source frame they form, and out to what
redshifts, as well as details of the physical conditions in the burst.

\acknowledgements This research has been supported through NASA
NAG5-13286 and the Monell Foundation. We are grateful to the referee
for useful comments, to X. Dai for assistance with {\it XSPEC} and
UNIX shell programming, and G. Chartas, B. Zhang, and N. Brandt for
helpful discussions.

 \end{document}